\documentclass[11pt,a4paper]{article}
\pdfoutput=1
\usepackage{jcappub}
\usepackage{bm}
\usepackage{indentfirst}
\usepackage{amsmath}
\usepackage{graphicx}
\usepackage{float}
\usepackage{amssymb}
\usepackage{subfigure}
\usepackage{amssymb}
\usepackage{psfrag}
\usepackage{hyperref}
\usepackage{tensor}
\usepackage{multirow}
\usepackage{color}
\usepackage{comment}
\usepackage{CJK}
\usepackage{array}
\usepackage{indentfirst}

\begin{document}



\newcommand{\tn}{\textnormal}	
	
\title{Detecting dark matter with extreme mass-ratio inspirals}

\author[a]{Chao Zhang,}
\author[a,1]{Guoyang Fu,\note{Corresponding author.}}
\author[b]{Ning Dai}

\affiliation[a]{School of Physics and Astronomy, Shanghai Jiao Tong University, 800 Dongchuan Rd, Shanghai 200240, China}
\affiliation[b]{School of Physics, Huazhong University of Science and Technology, Wuhan, Hubei 430074, China}

\emailAdd{zhangchao666@sjtu.edu.cn}
\emailAdd{fuguoyangedu@sjtu.edu.cn}
\emailAdd{daining@hust.edu.cn}


\abstract{Extreme mass ratio inspirals (EMRIs), where a small compact object inspiralls onto a supermassive black hole, are excellent sources for the space-based laser interferometer gravitational wave (GW) detectors.
The presence of dark matter surrounding the supermassive black hole will influence the binary orbital evolution and emitted gravitational waveform.
By direct observation of GW signals, we assess the detector's capability to detect whether an EMRI is immersed in a dark matter halo and to measure its characteristic spatial scale $a_0$ and mass $M_{\rm halo}$.
Apart from the GW emission, the dynamical friction and accretion caused by the dark matter halo can also affect the dynamics of an EMRI, leaving detectable signatures in the emitted gravitational signal.
We perform a Fisher-matrix error analysis to estimate the errors of parameters $a_0$ and $M_{\rm halo}$, as well as their correlation.
The results show that the highly correlated parameters $a_0$ and $M_{\rm halo}$ deteriorate the detector's ability to measure dark matter even though the dephasing and mismatch between signals with and without dark matter indicate much difference.
The effects of the dynamical friction and accretion can break possible degeneracies between parameters $a_0$ and $M_{\rm halo}$, thus greatly decreasing the uncertainty by about one order of magnitude.}

\maketitle

\section{Introduction}
The existence of dark matter (DM) is indirectly supported by astrophysical and cosmological observations over various ranges of scales, such as the rotation curves \cite{Rubin:1970zza,Begeman:1991iy}, bullet clusters \cite{Zwicky:1933gu,Clowe:2006eq}, large-scale structures \cite{Dietrich:2012mp}, and cosmic microwave background anisotropics \cite{Planck:2018vyg}.
However, the distribution and nature of DM remain unknown \cite{Bertone:2004pz,Bertone:2016nfn}.
The direct detection of gravitational waves (GWs) \cite{Abbott:2016blz,TheLIGOScientific:2017qsa,GBM:2017lvd,LIGOScientific:2018mvr,LIGOScientific:2020ibl,KAGRA:2021vkt} has opened new opportunities for offering more pronounced probes into the distribution and nature of DM \cite{Barausse:2020rsu,Macedo:2013qea,Eda:2013gg}.
It's known that a supermassive black hole (SMBH) of mass from $10^6~M_{\odot}$ to $10^{10}~M_{\odot}$ usually exists at the center of the galaxy, surrounded by a dark matter halo that is substantially larger than the visible galaxy \cite{Kormendy:2013dxa, Harris:2015vxa}.
The presence of the DM environment around SMBH may significantly alter the dynamics of a stellar-mass compact object with mass $m_p\sim1-100~M_{\odot}$ such as black holes (BHs), neutron stars, white dwarfs, etc. orbiting around SMBH.
Such binary systems, called extreme mass-ratio inspirals (EMRIs) \cite{Amaro-Seoane:2007osp,Babak:2017tow}, radiate millihertz GWs expected to be observed by the future space-based GW detectors like the Laser Interferometer Space Antenna (LISA) \cite{Danzmann:1997hm,LISA:2017pwj}, TianQin \cite{Luo:2015ght} and Taiji \cite{Hu:2017mde}.

There are various space-times used to match an isolated BH with a certain density distribution of DM halo.
The SMBH at the center of a galaxy on the certain distribution of DM halo was first studied by using Newtonian gravity \cite{Gondolo:1999ef}.
Recently, Cardoso et al \cite{Cardoso:2021wlq, Cardoso:2022whc} worked out the spacetime geometry with a spherically symmetric, static, non-vacuum BH generated by the DM distribution with the Hernquist density profile.
Unlike most studies worked on Newtonian inspiraled corrections, or computed leading quadrupolar GW energy flux to assess the contributions from nonvacuum background caused by DM, the generic formalism to calculate GW fluxes by EMRIs in the full relativistic region including the coupling between DM fluid and gravitational perturbations are established in \cite{Cardoso:2022whc,Figueiredo:2023gas}.
The evolution of EMRIs is not only affected by the GW energy fluxes but also relies on two mechanisms: dynamical friction and accretion.
When the stellar-mass compact object passes through a collisionless medium DM halo, the gravitational pull from the wake of the object slows it down, this effect is called dynamical friction \cite{Chandrasekhar:1943ys, Kim:2007zb, Zou:2022wtp}.
Besides, the object moving inside the DM halo will accrete the surrounding DM halo, thus slowing down, this effect is called accretion.
The accretion rate can be described by the Bondi-Hoyle-Lyttleton accretion \cite{Edgar:2004mk}.

These two mechanisms have already been explored in how they affect the orbits of compact object binaries and the emitted GWs \cite{Becker:2022wlo, Cardoso:2019rou, Kavanagh:2020cfn, Eda:2014kra, Li:2021pxf, Dai:2021olt}.
They would manifest as a dephasing in phase of the GWs emitted by the binary when compared to an equivalent system with/without accretion and dynamical friction, and also as a gradual change in GW amplitude.
However, previous works were mainly studied on Newtonian inspiraled corrections and needed to be extended to the full relativistic zone.
As discussed by \cite{Cardoso:2022whc}, to linear order in $M_{\rm halo}/a_0$, axial perturbations do not couple to DM perturbations, and we can use a simple propagation redshift to describe the result.
However, polar fluctuations are coupled to the DM fluid and less prone to redshift effects so a naive redshift is not sufficient to describe GW generation and propagation, even in regions of parameter space corresponding to large, near-galactic scales \cite{Speeney:2024mas}.
Thus, polar fluctuations together with the axial sector can break possible degeneracies, with sensitive, low-frequency detectors.
Obviously, dynamical friction and accretion are coupled to the DM fluid so it remains a question whether they can break possible degeneracies deeper to decrease the uncertainty of DM parameters $M_{\rm halo}$ and $a_0$.

In this paper, we consider the EMRI system that a stellar-mass BH inspirals into an SMBH surrounded by DM halo.
The evolution of the orbit and the waveforms are obtained by implementing the BH perturbation method established by \cite{Cardoso:2022whc}.
We perform the Fisher information matrix to estimate errors of DM parameters such as DM mass $M_{\rm halo}$ and the typical length scale of a galaxy or halo radius $a_0$ taking into account the accretion and dynamical friction.
We take LISA as a representative of the space-based detectors in our discussion,
and the analysis can be easily extended to other space-based detectors.
The paper is organized as follows.
In Sec. \ref{sec2}, we introduce the formalism to study GW emission by EMRIs in spherically symmetric, non-vacuum black-hole spacetimes.
In Sec. \ref{sec3}, we numerically calculate the energy GW fluxes in the background of an SMBH surrounded by DM halo and study the orbital evolution under the influence of accretion and dynamical friction.
In Sec. \ref{sec4}, we calculate the mismatch between GWs with and without DM and perform the FIM to estimate the errors of detecting DM mass $M_{\rm halo}$ and its typical length scale $a_0$ with LISA.
Sec. \ref{sec5} is devoted to conclusions and discussions.
In this paper, we set $G=c=1$.

\section{Dark matter profile and perturbation equations}\label{sec2}
A good description of galactic profiles guided by observations and large-scale situations is a Hernquist-type density distribution \cite{Hernquist:1990be}.
A semi-analytical model can describe Hernquist-type density distribution
\begin{equation}
\rho(r)=\rho_0 (r/a_0)^{-\gamma}[1+(r/a_0)^\alpha]^{(\gamma-\beta)/\alpha}\ ,
\label{math:densityp}
\end{equation}
where $\beta=4$ and $\gamma=1$ determine
the dependence of the profile at large and small
distances, respectively, with the sharpness of the
transition given by $\alpha=1$ \cite{Hernquist:1990be,Graham:2005xx}.
The slope of the distribution changes at a characteristic
spatial scale determined by $a_0$, with $\rho_0$ being
the corresponding density.
We briefly summarize the methods in \cite{Figueiredo:2023gas,Cardoso:2021wlq,Cardoso:2022whc,Cardoso:2020iji,Einstein:1939ms} to calculate a static, spherically
symmetric BH spacetime embedded in an environment with a generic density profile $\rho(r)$.
The background metric specified by the line element
\begin{equation}\label{background}
ds^2=g_{\mu\nu}^{(0)}dx^\mu dx^\nu=-a(r) dt^2 + \frac{d r^2}{1-\frac{2m(r)}{r}} + r^2 d\Omega^2 \ ,
\end{equation}
is a solution of the sourced Einstein's field equations $G^{(0)}_{\mu\nu}=8\pi T^\tn{(0)env}_{\mu\nu}$ and the anisotropic stress-energy
tensor has the following form:
\begin{equation}
    (T^\tn{(0)env})^{\mu}{_\nu} = {\rm{diag}}(-\rho,0,P,P)\ .
\end{equation}
For a given choice of $\rho(r)$, the background metric solution is completely determined by
\begin{equation}
\frac{a'(r)}{a(r)}=\frac{2m(r)/r}{r-2m(r)}\quad \ ,\quad
P(r) =\frac{m(r)/2}{r-2m(r)}\rho(r)\ ,\label{math:back}
\end{equation}
where a prime denotes a derivative with respect to the radial coordinate.
To solve the background metric, we start by choosing a density profile $\rho(r)$ according to the Hernquist model and then integrate the equation for the mass
function $m(r)$ from the horizon $r_\tn{h}=2~M_\tn{BH}$,
where $m(r_h)=M_\tn{BH}$.
Using the above ansatz for the static, spherically symmetric spacetime, we can get the background metric \cite{Cardoso:2021wlq}
\begin{equation}
\begin{split}
\rho(r)&=\frac{M_{\rm halo}(a_0+2M_{\rm BH})(r-2M_{\rm BH})}{2\pi r^2(r+a_0)^3},\\
    m(r)&=M_{\rm BH}+\frac{M_{\rm halo} r^2}{(a_0+r)^2}\left(1-\frac{2M_{\rm BH}}{r}\right)^2,\\
    a(r)&=\left(1-\frac{2M_{\rm BH}}{r}\right)e^\Upsilon,\\
    \Upsilon&=-\pi\sqrt{\frac{M_{\rm halo}}{\xi}}+2\sqrt{\frac{M_{\rm halo}}{\xi}}\arctan\left(\frac{r+a_0-M_{\rm halo}}{\sqrt{M_{\rm halo}\xi}}\right),\\
    \xi&=2a_0-M_{\rm halo}+4M_{\rm BH}.
\end{split}
\end{equation}
Note that $M_\tn{halo}$ is the total environmental mass outside the SMBH and  $a_0$ is DM characteristic spatial scale.
The solution for $m(r)$ and $a(r)$ allows us to compute the tangential pressure $P(r)$, as well as the geodesic quantities, like the circular orbit frequency $\Omega$ and the conserved energy quantity $E$ per unit mass
\begin{equation}
    \Omega=\left[\frac{1}{r^2} \frac{a(r) m(r)}{r-2m(r)} \right]^{1/2}\ ,~
    E=\left[\frac{r-2m(r)}{r-3m(r)}a(r)\right]^{1/2}\ .
\end{equation}
and the innermost circular orbit (ISCO) $r_\tn{ISCO}$ is given by a solution of the following
equation
\begin{equation}
r^2m'(r) + rm(r) - 6m^2(r)=0.
\end{equation}
We consider the stellar-mass compact object as a perturbed to the background spacetime in Eq. \eqref{background} with the forms
\begin{equation}
      g_{\mu\nu} = g_{\mu\nu}^{(0)} + g_{\mu\nu}^{(1)},~~~ T^{\rm{DM}}_{\mu\nu} = T^{\rm{DM}(0)}_{\mu\nu} + T^{\rm{DM}(1)}_{\mu\nu},  
\end{equation}
where superscript (1) denotes the perturbations and `DM' denotes the dark matter.
In the Regge-Wheeler-Zerilli gauge  \cite{Zerilli:1970se, Zerilli:1970wzz}, the metric perturbation can be decomposed into tensor spherical harmonics as $g_{\mu\nu}^{(1)} = g_{\mu\nu}^{(1) \textrm{axial}}+g_{\mu\nu}^{(1) \textrm{polar}}$, where $g_{\mu\nu}^{(1) \textrm{axial}}$ and  $g_{\mu\nu}^{(1) \textrm{polar}}$ represent the axial and polar perturbations respectively,
\begin{equation} \label{prt1}
    \begin{aligned}
    g_{\mu\nu}^{(1) \textrm{axial}} =& \sum_{l=2}^{\infty}\sum_{m=-l}^{l} \frac{\sqrt{2l(l+1)}}{r}\Big(ih_{1}^{lm}c_{lm,\mu\nu}-h_{0}^{lm}c^{0}_{lm, \mu\nu}\Big),  \\
    g_{\mu\nu}^{(1) \textrm{polar}} =& \sum_{l=2}^{\infty}\sum_{m=-l}^{l}\Big(aH_{0}^{lm}a^{0}_{lm,\mu\nu}-i\sqrt{2}H_{1}^{lm}a^{1}_{lm,\mu\nu}+\frac{H_{2}^{lm}}{(1-2m/r)}a_{lm,\mu\nu}+\sqrt{2}K^{lm}g_{lm,\mu\nu}\Big).
\end{aligned}
\end{equation}
Here, $\{c_{lm,\mu\nu}, c^{0}_{lm,\mu\nu}, a^{0}_{lm,\mu\nu}, a^{1}_{lm,\mu\nu}, a_{lm,\mu\nu}, g_{lm,\mu\nu}\}$ are six tensor spherical harmonics in \cite{Sago:2002fe}.
The perturbed density, pressure of the dark matter halo can also be decomposed into tensor spherical harmonics in the following manner
\begin{equation}
\begin{split}
\rho^{(1)}(t,r,\theta,\phi)&=\sum_{l=2}^\infty\sum_{m=-l}^{l}\delta \rho^{l m}(t,r)Y_{l m}(\theta,\phi), \ \\
p^{(1)}_t(t,r,\theta,\phi)&=\sum_{l=2}^\infty\sum_{m=-l}^{l}\delta p_{t}^{l m}(t,r)Y_{l m}(\theta,\phi),  \\
p^{(1)}_r(t,r,\theta,\phi)&=\sum_{l=2}^\infty\sum_{m=-l}^{l}\delta p_{r}^{l m}(t,r)Y_{l m}(\theta,\phi),
\end{split}
\end{equation}
where $Y_{lm}(\theta, \phi)$ denotes the spherical harmonics on 2-sphere.
A barotropic equation of state provides a further relation between pressure, density variations and the medium's speed of sound via
\begin{equation}
\delta p_{t,r}^{l m}(t,r)  =c_{s_{t,r}}^2(r) \delta \rho^{lm}(t,r),
\end{equation}
where $c_{s_r}$ and $c_{s_t}$ are the radial and transverse sound speeds, respectively.
We choose the values with $c_{s_{r,t}}=(0.9,0)$ throughout the paper.
The perturbed energy-momentum tensor of the environment $T^{\textrm{DM}(1)}_{\mu\nu}$ can be constructed in \cite{Cardoso:2022whc}.
The perturbed field equation is given by
\begin{align}
    G^{(1)}_{\mu\nu} - 8\pi (T_{\mu\nu}^{\textrm{DM} (1)}+T^{P}_{\mu\nu})=0,
\end{align}
where $G^{(1)}_{\mu\nu}$ is the perturbed Einstein tensor and $T^{P}_{\mu\nu}$ is energy-momentum tensor of the stellar-mass compact object
\begin{align}\label{src}
    T^{\mu\nu}_{P} = \mu \int d\tau \frac{\delta^{(4)}(x^{\mu}-z^{\mu}_{P}(\tau))}{\sqrt{-g}}U^{\mu}_{P}U^{\nu}_{P},
\end{align}
where $\mu$ is the mass of the secondary object and $U^{\mu}_{P}$ denotes the four-velocity. $\tau$ is the proper time along the worldline $z^{\mu}_{P}$ and $U^{\mu}_{P} = dz^{\mu}_{P}/d\tau$ represents the tangent to the line.
In the frequency domain, the axial perturbed field equation in terms of a master variable $\chi_{lm\omega}=h_1^{lm}\sqrt{a(1-2m/r)}/r$ is given by
\begin{align}
    \Big(\frac{d^{2}}{dr_{*}^{2}}+\omega^{2}-V^{ax}\Big)\chi_{lm\omega} = S^{ax}_{lm\omega}, \label{eq:master_axial}
\end{align}
where the tortoise coordinate follows the relation $dr_{*}=dr/\sqrt{a(1-2m/r)}$, and the potential is given by
\begin{equation}
V^{ax} = \dfrac{a}{r^2}\Big(l(l+1) - \dfrac{6m(r)}{r} + m'(r)  \Big).
\end{equation}
The polar perturbed field equation is given by
\begin{align}\label{ppte1}
  \frac{d\vec{\psi}_{lm\omega}}{dr}-\boldsymbol{\alpha}\vec{\psi}_{lm\omega}=\vec S^{pol}_{lm\omega} ,
\end{align}
where $\vec{\psi}_{lm\omega}=(H_1^{lm}, H_0^{lm}, K^{lm}, W^{lm},\delta\rho^{lm})$.
The components of $S^{ax}_{lm\omega}$, $\vec S^{pol}_{lm\omega}$, $\boldsymbol{\alpha}$, $W^{lm}$ and $H^{lm}_2$  are given in \cite{Cardoso:2022whc}.
With the solution of the axial and polar perturbation equation, we can calculate the energy flux at infinity using the following relation \cite{Thorne:1980ru, Poisson:2004cw, Martel:2003jj}
\begin{equation}\label{GWdotE}
\left\langle \frac{d E}{dt} \right\rangle_\text{GW}=\sum_{lm}\frac{dE_{lm}}{dt}=\frac{1}{32\pi}\sum_{lm}\frac{(l+2)!}{(l-2)!}\left[|\dot Z_{lm\omega}^{\textrm{polar}}|^2+4|Z_{lm\omega}^{\textrm{axial}}|^2\right]\,,\\
\end{equation}
where
\begin{equation}
\begin{split}
Z_{lm\omega}^{\textrm{axial}}&=\chi_{lm\omega},\\
Z_{lm\omega}^{\textrm{polar}}&=\frac{r}{n+1}\left[K^{lm}+\frac{a}{n}\left(H^{lm}_2-r\frac{\partial K^{lm}}{\partial r}\right)\right],
\end{split}
\end{equation}
and $n=l(l+1)/2-1$.
We utilize the packages \cite{GRIT_REPO, SGREP_REPO} to obtain the GW energy flux emitted from the second object inspiralling around SMBH surrounded DM in a circular equatorial trajectory.

\section{Orbital evolution}\label{sec3}
\begin{figure}
    \centering
    \includegraphics[width=0.9\columnwidth]{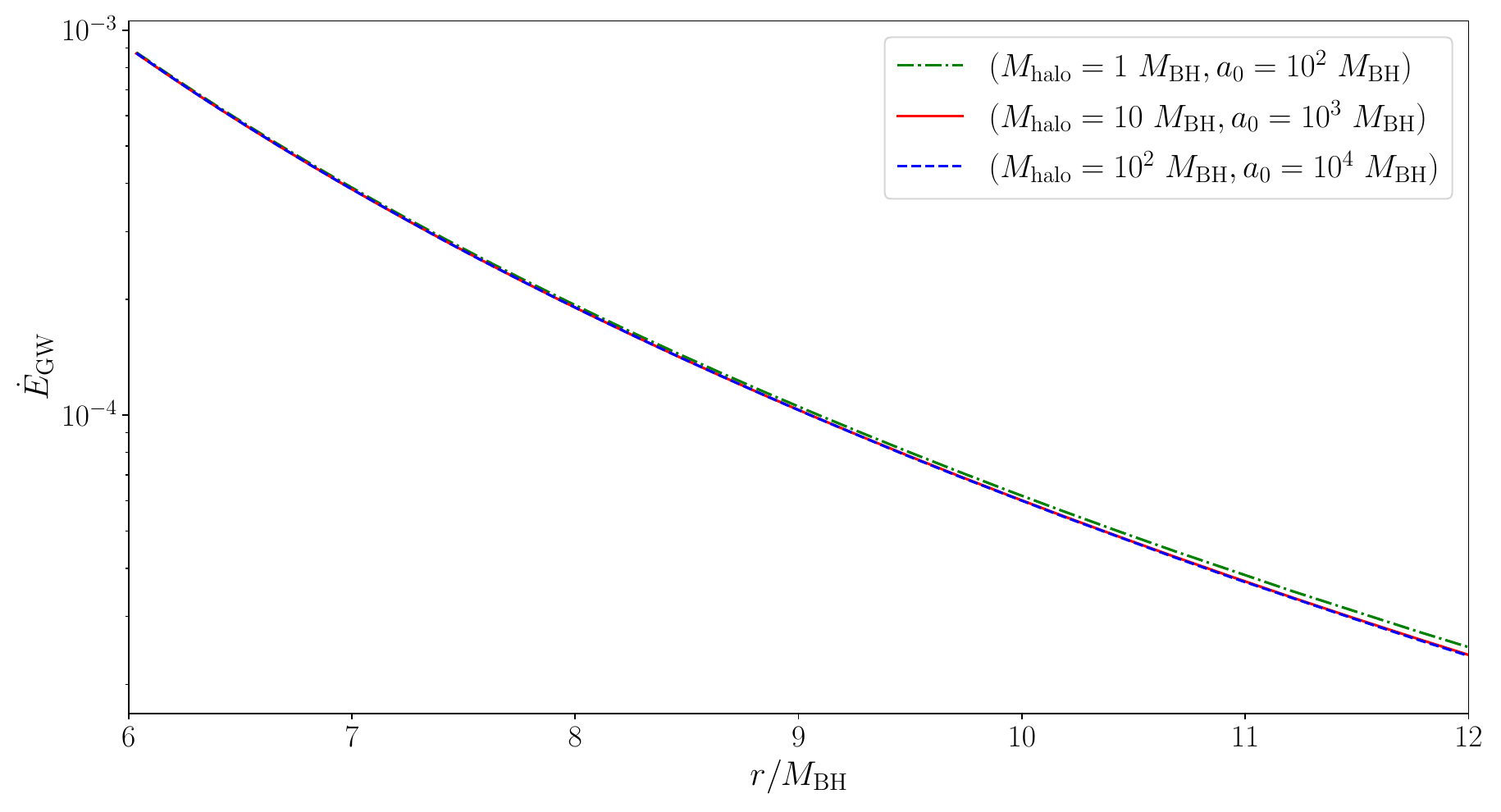}
    \caption{The total energy fluxes versus the orbital distance for given dark matter configuration.}
    \label{energyflux}
\end{figure}

As seen in Eq.~\eqref{GWdotE}, the numerical calculation of energy flux involves the truncation of infinite sums over $l$ and $m$.
This truncation limits the accuracy of the numerical results.
In Table~\ref{comparision} we compare our results on the radiated energy flux of different modes with \cite{Cardoso:2022whc} for three representative points.
 \begin{table}[t!]\centering
	\begin{tabular}{c c c c |c c c c c c}
		\hline\hline
		$l$ & \hspace{3mm} $m$ & \hspace{3mm} $r$ & \hspace{3mm} $\langle\dot{E}_{lm}\rangle$ & \hspace{3mm}		$l$ & \hspace{3mm} $m$ & \hspace{3mm} $r$ & \hspace{3mm} $\langle\dot{E}_{lm}\rangle$ \\
		\hline\hline
		\multirow{3}*{2} & \hspace{3mm}\multirow{3}*{1} & \hspace{3mm} 6.1 & \hspace{3mm} 3.68932e-6 \hspace{2mm} & \multirow{3}*{2} & \hspace{3mm}\multirow{3}*{2} & \hspace{3mm} 6.1 & \hspace{3mm} 5.88689e-4  \\
						 &	      & \hspace{3mm} 7.9456 & \hspace{3mm} 6.91578e-7  	 \hspace{2mm}&	  	&		    & \hspace{3mm} 7.9456 & \hspace{3mm}  1.62097e-4    \\
       				 &	  			    & \hspace{3mm} 11 & \hspace{3mm} 1.00565e-7 	\hspace{2mm} &	  	&		    & \hspace{3mm} 11 & \hspace{3mm} 3.95057e-5  \\ \hline
        \multirow{3}*{3} & \hspace{3mm}\multirow{3}*{1} & \hspace{3mm} 6.1 & \hspace{3mm} 6.08585e-9 \hspace{2mm} & \multirow{3}*{3} & \hspace{3mm}\multirow{3}*{3} & \hspace{3mm} 6.1 & \hspace{3mm} 1.13009e-4  \\
						 &	      & \hspace{3mm} 7.9456 & \hspace{3mm}   9.64039e-10	 \hspace{2mm}&	  	&		    & \hspace{3mm} 7.9456 & \hspace{3mm}  2.34893e-5    \\
       				 &	  			    & \hspace{3mm} 11 & \hspace{3mm} 1.66321e-11 	\hspace{2mm} &	  	&		    & \hspace{3mm} 11 & \hspace{3mm}  4.00239e-6  \\ \hline
        \multirow{3}*{3} & \hspace{3mm}\multirow{3}*{2} & \hspace{3mm} 6.1 & \hspace{3mm} 1.49177e-6 \hspace{2mm} & \multirow{3}*{4} & \hspace{3mm}\multirow{3}*{4} & \hspace{3mm} 6.1 & \hspace{3mm} 2.68113e-5  \\
						 &	      & \hspace{3mm} 7.9456 & \hspace{3mm}   2.16117e-7	 \hspace{2mm}&	  	&		    & \hspace{3mm} 7.9456 & \hspace{3mm}  4.08356e-6   \\
       				 &	  			    & \hspace{3mm} 11 & \hspace{3mm}  2.34463e-8	\hspace{2mm} &	  	&		    & \hspace{3mm} 11 & \hspace{3mm} 4.33964e-7  \\ \hline
        \multirow{3}*{4} & \hspace{3mm}\multirow{3}*{3} & \hspace{3mm} 6.1 & \hspace{3mm} 4.52640e-7 \hspace{2mm} & \multirow{3}*{5} & \hspace{3mm}\multirow{3}*{5} & \hspace{3mm} 6.1 & \hspace{3mm} 7.02725e-6  \\
						 &	      & \hspace{3mm} 7.9456 & \hspace{3mm}   5.02517e-8 	 \hspace{2mm}&	  	&		    & \hspace{3mm} 7.9456 & \hspace{3mm}  8.09661e-7  \\
       				 &	  			    & \hspace{3mm} 11 & \hspace{3mm} 4.04301e-9	\hspace{2mm} &	  	&		    & \hspace{3mm} 11 & \hspace{3mm} 6.07953e-8 \\
		\hline\hline
	\end{tabular}
\caption{
Average energy flux $\langle\dot{E}_{lm}\rangle$ for different modes at different semi-latus radius points. We provide the data considering $M_{\rm halo}=10~M_{\textrm{BH}}$ and $a_0=100~M_{\textrm{BH}}$. Polar modes are represented with even values of $l+m$ modes, whereas odd values of $l+m$ correspond to axial modes.
}\label{comparision}
\end{table}
In order to reach the accuracy goal about $10^{-2}$ \cite{Cutler:1994pb}, we can only choose $(l,m)=(2,2)$, $(2,1)$, $(3,2)$, $(3,3)$ and $(4,4)$ modes.
Figure \ref{energyflux} shows the normalized GW energy flux $\mu^{-2}M_{\rm BH}^2\dot{E}_{\rm GW}$ for a particle on a circular orbit about a SMBH surrounded DM halo, as a function of orbital radius.
We find that the values of the energy flux are not only sensitive to the effects of a specific parameter termed as halo compactness $(M_{\rm halo}/a_0)$ that ascertains the gravitational features or properties of a galaxy but also to the parameter $(M_{\rm halo}/a_0^2)$ that describes the density of DM.
For the same halo compactness $M_{\rm halo}/a_0=10^{-2}$, the energy flux decreases as the density $(M_{\rm halo}/a_0^2)$ decreases and finally reaches almost the constant value.
As discussed by \cite{Cardoso:2022whc}, a naive redshift is not sufficient to describe GW generation and propagation.
The total energy flux consisting of the polar and axial sectors can break possible degeneracies to improve the ability to detect DM parameters $M_{\rm halo}$ and $a_0$.

\begin{figure}
    \centering
    \includegraphics[width=0.98\columnwidth]{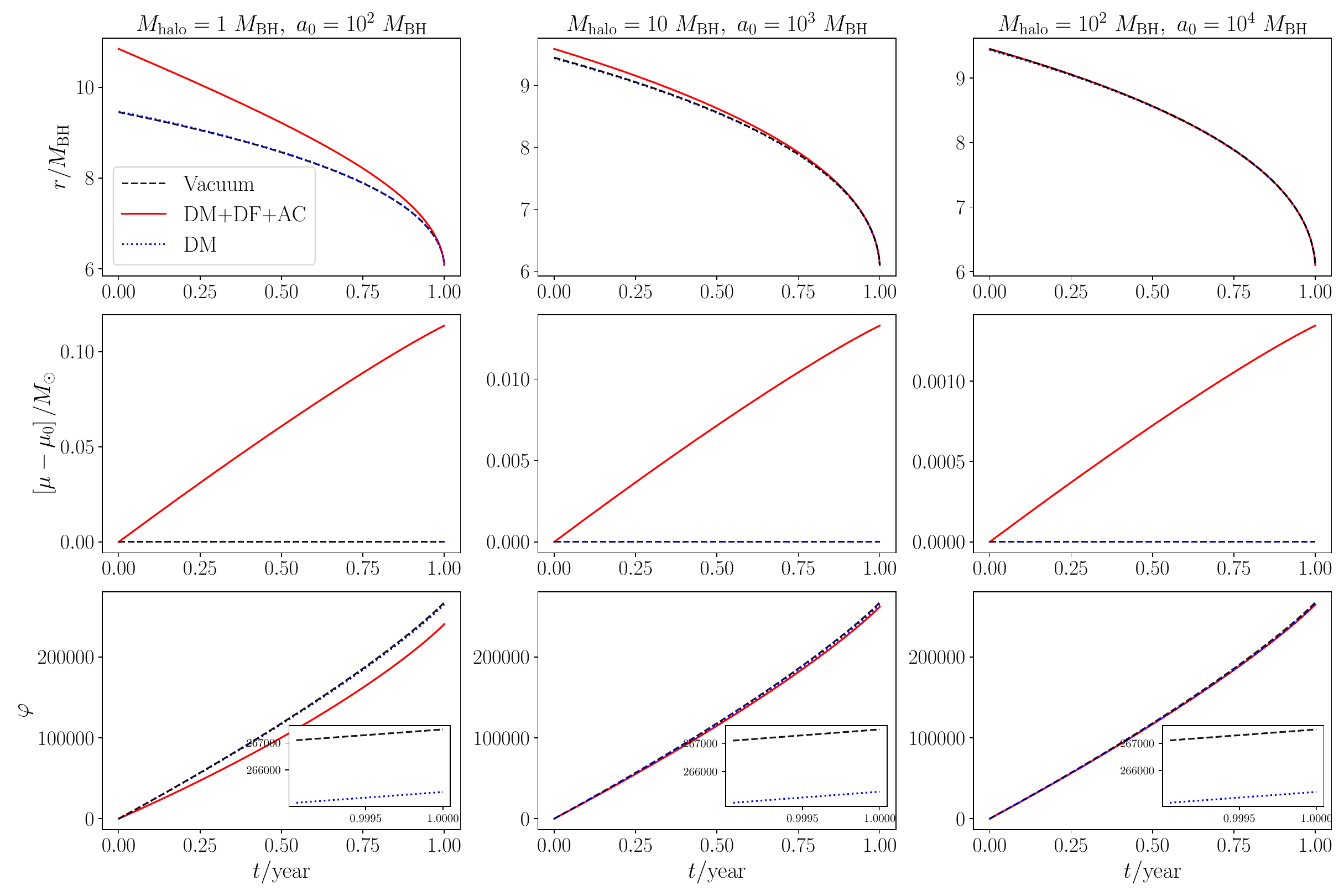}
    \caption{The orbital evolution for different configurations with $M_{\rm halo}$ and $a_0$. 
    The initial orbital separation for the vacuum case is $r_0=9.45022$.
    The initial orbital separation is $r_0$= $10.8443$, $9.59183$ and $9.45394$ with dynamical friction and accretion for specific configuration with $(M_{\rm halo}, a_0)=(1,10^2), (10,10^3), (10^2,10^3)$, respectively.
    The initial orbital separation is $r_0$= $9.46618$, $9.43950$ and $9.43864$ without dynamical friction and accretion for specific configuration with $(M_{\rm halo}, a_0)=(1,10^2), (10,10^3), (10^2,10^3)$, respectively.  
  The top panels show the evolution of the orbital separation $r$ with time, the middle panels show the evolution of the compact mass of $\mu$ with time, and the bottom panels show the evolution of the orbital phase $\varphi$ with time.
  "$\rm DM+DF+AC$" means the effect including dynamical friction and accretion."$\rm DM$" means the effect without dynamical friction and accretion.}
    \label{orbit}
\end{figure}

When the object with mass $\mu$ passes through a collisionless medium DM, the object is decelerated in the direction and loses its kinetic energy because of dynamical friction and accretion.
The dynamical friction is \cite{Kim:2007zb,Zou:2022wtp}
\begin{equation}
\label{eq:F_DF}
\left\langle \frac{d E}{dt} \right\rangle_{\rm DF}=-\frac{4\pi\rho \mu^2}{v}I_\phi(v_{cs}),
\end{equation}
and 
\begin{equation}
I_\phi=\begin{cases}
		v_{cs}^3/3 - 0.80352 v_{cs}^4 + 7.68585 v_{cs}^5,& \text{if }v_{cs}<0.08588, \\
	0.7706\ln\left(\frac{1+v_{cs}}{1.0004-0.9185v_{cs}}\right)-1.4703v_{cs},&
		\text{if } 0.08588<v_{cs}<1.0, \\
	\ln[3300(v_{cs}-0.71)^{5.72}v_{cs}^{-9.58}],&
		  \text{if } 1.0\le v_{cs}<4.4, \\
		\ln\left(\frac{10}{0.11v_{cs}+1.65}\right),&
		\text{if } v_{cs}\ge 4.4,
	\end{cases}
\end{equation}
where $v_{cs}=v/c_s$, $v=\Omega r$ is the speed of the object, $c_s=\sqrt{dP/d\rho}$ is the sound speed.
The accretion of the object is \cite{Edgar:2004mk},
\begin{equation}\label{eq:F_AC}
\left\langle \frac{d E}{dt} \right\rangle_{\rm AC}=-\frac{4\pi\rho\,\mu^2}{(c_s^2+v^2)^{3/2}}v^2,
\end{equation}
where
\begin{equation}
\label{eq:mDdot}
\dot{\mu}=\frac{4\pi\rho\,\mu^2}{(c_s^2+v^2)^{3/2}}
\end{equation}
is the accretion rate, an overdot indicates the differentiation with respect to time.
The evolution of the orbital radius $r$ is determined by the energy balance condition,
\begin{equation}\label{eq:drdt}
\begin{split}
	\frac{dE}{dt}
	=\left\langle \frac{d E}{dt} \right\rangle_\text{GW}+\left\langle \frac{d E}{dt} \right\rangle_{\rm DF}+\left\langle \frac{d E}{dt} \right\rangle_{\rm AC}.
\end{split}
\end{equation}
Combining Eqs. \eqref{eq:F_DF}-\eqref{eq:drdt}, we can obtain the evolution of the orbital radius $r(t)$ and the object mass $\mu(t)$.
With $r(t)$, the orbital phase is determined as
\begin{equation}
\label{eq:omega}
\varphi(t)=\varphi_0+\int_0^t\Omega(r(t))dt,
\end{equation}
where $\varphi_0$ and $\mu_0$ is the initial orbital phase and object mass at $t=0$, respectively.
In Fig.~\ref{orbit}, we show the evolution of the semi-latus rectum, compact mass, and the orbital phase for different values of dark matter parameters and with/without dynamical friction and accretion.
We set the observation time as one year.
The initial orbital separation $r_0$ is adjusted to experience the one-year adiabatic evolution before the ISCO.
To make a comparison with the vacuum scenario, we also show the evolution of the orbital parameters for the Schwarzschild BH.
We can see from Fig.~\ref{orbit} that the initial orbital separation $r_0$ is smaller or larger in the presence of dark matter without dynamical friction and accretion but always larger with dynamical friction and accretion, compared with the vacuum scenario, which means that the effect of DM gravitational features can slow down or accelerate the decrease of the semi-latus rectum depending on specific configuration while dynamical friction and accretion accelerate the decrease of the semi-latus rectum.
The total energy flux emitted from EMRI surrounded by DM depending on the specific DM parameters $M_{\rm halo}$ and $a_0$ may be larger or smaller than the vacuum case \cite{Speeney:2024mas}.
So the effect of DM gravitational features depends on the specific DM parameters.
However, dynamical friction and accretion always accelerate the orbital evolution.
Taking into account dynamical friction and accretion, the mass of the compact object increases during the orbital evolution and depends on the density of DM.
The larger the density of DM, the stronger the effect of dynamical friction and accretion.
The dephasing of orbit with and without DM is far larger than the threshold for a detectable dephasing that two signals are distinguishable by space-based GW detector as 1 \cite{Berti:2004bd,Maselli:2020zgv}.

\section{Results}\label{sec4}
\begin{figure}
    \centering
    \includegraphics[width=0.8\columnwidth]{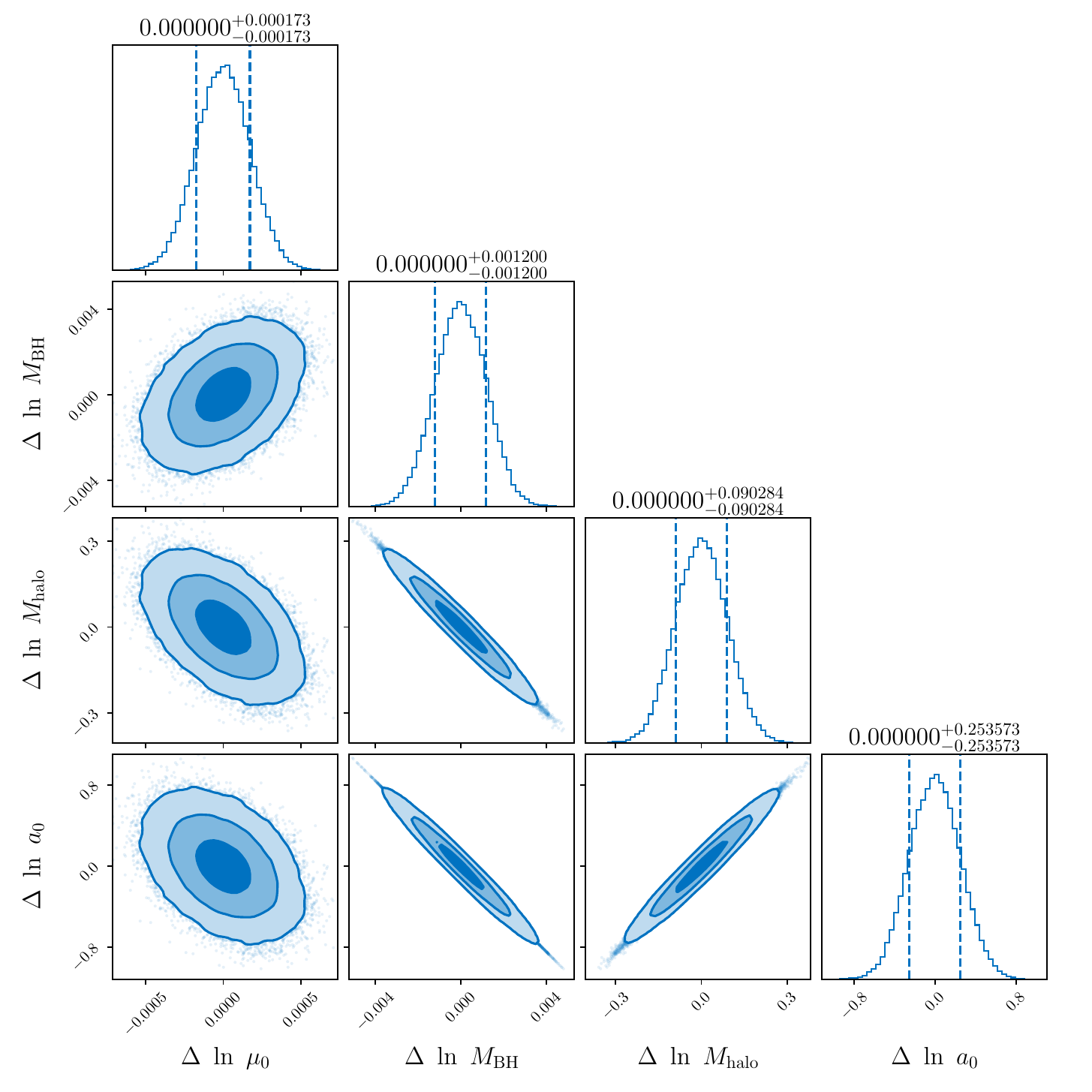}
    \caption{Corner plot for the probability distribution of masses and DM parameters, inferred after EMRI with $(M_{\rm halo}=1~M_{\rm BH},a_0=100~M_{\rm BH})$ and taking no into account dynamical friction and accretion. Diagonal boxes refer to marginalized distributions. Vertical lines show the 1-$\sigma$ interval for each waveform parameter. The contours correspond to the $68\%$, $95\%$, and $99\%$ probability confidence intervals.}
    \label{DMcorner}
\end{figure}

To make the analysis more accurate and account for the degeneracy among parameters,
we calculate the mismatch between two GW waveforms and carry out parameter estimation with the Fisher information matrix (FIM) method.
Based on previous analysis \cite{Piovano:2021iwv,Zhang:2023ceh}, higher harmonics can only greatly influence the exterior parameters and have nearly no influence on the intrinsic parameters.
We can only compute the GW signal in the quadrupole approximation \cite{Barack:2003fp,Huerta:2011kt,Jiang:2021htl}.
The GW waveform from EMRIs is given by
\begin{equation}
h_+=\frac{4\Omega^2 r^2}{d_L}\frac{1+\cos^2\iota}{2}\cos\left[2\varphi(t)\right],
\end{equation}
\begin{equation}
h_\times=\frac{4\Omega^2 r^2}{d_L}\cos\iota\sin\left[2\varphi(t)\right],
\end{equation}
where $d_L$ is the distance to the source, $r$ is the orbital radius, and $\iota$ is the inclination angle, which is the angle between the line-of-sight and the rotational axis of the orbits.
The GW strain measured by the detector is
\begin{equation}\label{signal}
h(t)=h_{+}(t) F^{+}(t)+h_{\times}(t) F^{\times}(t),
\end{equation}
where the interferometer pattern functions $F^{+,\times}(t)$ and $\iota$ can be expressed in terms of the source orientation $(\theta_s,\phi_s)$ and the direction of the angular momentum $(\theta_1,\phi_1)$.
Finally, the GW signals are modulated due to the orbital motion \cite{Babak:2006uv}.
We account for this effect by modifying the phase as
\begin{equation}\label{doppler}
\varphi(t)\to \varphi(t)+\varphi'(t)R_{\text{AU}}\sin\theta_s\cos\left(\frac{2\pi t}{T} -\phi_s-\phi_{\alpha}\right),
\end{equation}
where $\phi_{\alpha}$ is the ecliptic longitude of the detector $\alpha$ at $t=0$,
the rotational period $T$ is 1 year and the radius of the orbit $R_{\text{AU}}$ is 1 AU.
The signal-to-noise ratio (SNR) of the GW signals is
\begin{equation}
\rho=\sqrt{\left\langle h|h \right\rangle},
\end{equation}
the noise-weighted inner product between two templates $h_1$ and $h_2$ is
\begin{equation}\label{product}
\left\langle h_{1} \mid h_{2}\right\rangle=4 \Re \int_{f_{\min }}^{f_{\max }} \frac{\tilde{h}_{1}(f) \tilde{h}_{2}^{*}(f)}{S_{n}(f)} df,
\end{equation}
where
\begin{equation}
    f_{\text{max}}=\text{min}(f_{\text{ISCO}},f_{\text{up}}),~~~~~~f_{\text{min}}=\text{max}(f_{\text{low}},f_{\text{start}}),
\end{equation}
$f_{\text{ISCO}}$ is the GW frequency at the ISCO,
$f_{\text{start}}$ is the initial frequency at $t=0$,
the cutoff frequencies $f_{\text{low}}=10^{-4}$ Hz and $f_{\text{up}}=1$ Hz, $\tilde{h}(f)$ is the Fourier transform of the time-domain signal $h(t)$,
its complex conjugate is $\tilde{h}^{*}(f)$,
and $S_n(f)$ is the noise spectral density of the space-based GW detectors \cite{Chua:2015mua, Katz:2021yft, Barsanti:2022ana}.
The mismatch between two signals is defined as
\begin{equation}\label{eq:def_F}
\rm {Mismatch}=1-\max_{\{t_c,\phi_c\}}\frac{\langle h_1\vert
	h_2\rangle}{\sqrt{\langle h_1\vert h_1\rangle\langle h_2\vert h_2\rangle}}\ ,
\end{equation}
where $(t_c,\phi_c)$ are time and phase offsets \cite{Lindblom:2008cm}.
Two waveforms are considered experimentally distinguishable if their mismatch is larger than $d/(2\rho^2)$ with $d$ being the number of parameters \cite{Jiang:2021htl}.
To consider degeneracies among the source parameters and give a more accurate analysis on the detectability of the dark matter with LISA,
we carry out the parameter estimation with the FIM method.
In the time domain, the GW signal is mainly determined by parameters
\begin{equation}
\xi=(\ln M_{\rm{BH}}, \ln \mu_0,\ln M_{\rm{halo}}, \ln a_0, r_0, \theta_s, \phi_s, \theta_1, \phi_1, d_L),
\end{equation}
where $\mu_0$ and $r_0$ is the mass and radius of the smaller compact object at $t=0$.
In the large SNR limit,
the covariances of source parameters $\xi$  are given by the inverse of the Fisher information matrix
\begin{equation}
\Gamma_{i j}=\left\langle\left.\frac{\partial h}{\partial \xi_{i}}\right| \frac{\partial h}{\partial \xi_{j}}\right\rangle_{\xi=\hat{\xi}}.
\end{equation}
The statistical error on $\xi$ and the correlation coefficients between the parameters are provided by the diagonal and non-diagonal parts of ${\bf \Sigma}={\bf \Gamma}^{-1}$, i.e.
\begin{equation}
\sigma_{i}=\Sigma_{i i}^{1 / 2} \quad, \quad c_{\xi_{i} \xi_{j}}=\Sigma_{i j} /\left(\sigma_{\xi_{i}} \sigma_{\xi_{j}}\right).
\end{equation}
Because of the triangle configuration of the space-based GW detector regarded as a network of two L-shaped detectors, with the second interferometer rotated of $60^\circ$ with respect to the first one, the total SNR can be written as the sum of SNRs of two L-shaped detectors \cite{Cutler:1994pb}
\begin{equation}
\rho=\sqrt{\rho_1^2+\rho_2^2}=\sqrt{\left\langle h_1|h_1 \right\rangle+\left\langle h_2|h_2 \right\rangle},
\end{equation}
where $h_1$ and $h_2$ denote the signals detected by two L-shaped detectors.
The total covariance matrix of the source parameters is obtained by inverting the sum of the Fisher matrices $\sigma_{\xi_i}^2=(\Gamma_1+\Gamma_2)^{-1}_{ii}$.
The source angles are fixed $\theta_s=\pi/3,~\phi_s=\pi/2$, $\theta_1=\pi/4,~\phi_1=\pi/4$, and the initial orbital separation $r_0$ is adjusted to experience one-year adiabatic evolution before the final plunge $r_{\text{end}}=r_{\text{ISCO}}+0.1~M_{\rm BH}$.
We consider the EMRI system with $\mu_0=10~M_{\odot}$, $M_{\rm BH}=10^6~M_{\odot}$, different DM parameters values of $M_{\rm{halo}}$ and $a_0$, and the luminosity distance $d_L$ can be changed freely to vary the SNR of the signal.
To analyze the effects of dynamical friction and accretion, we calculate the mismatch as well as FIM with and without dynamical friction and accretion.
The details can be seen in the Appendix \ref{FIMsetup}.
Figure \ref{DMACcorner} shows the probability distribution obtained
using the Fisher matrix approach for the component masses, DM parameters, $(\mu_0, M_{\rm BH}, M_{\rm halo},a_0)$, for EMRIs observed one year before ISCO with $(M_{\rm halo}=1~M_{\rm BH},a_0=100~M_{\rm BH})$ and SNR of 150 taking into account dynamical friction and accretion. 
We obtain the error for $\sigma_{\ln M_{\rm halo}}=3.7\times 10^{-3}$ and $\sigma_{\ln a_0}=6.1\times 10^{-3}$.
Figure \ref{DMcorner} shows the same result without regard to dynamical friction and accretion.
We obtain the error for $\sigma_{\ln M_{\rm halo}}=9.0\times 10^{-2}$ and $\sigma_{\ln a_0}=2.5\times 10^{-1}$.
We can see that including dynamical friction and accretion can break possible degeneracies deeper to improve the ability to detect DM parameters $M_{\rm halo}$ and $a_0$ about one order of magnitude.
\begin{figure}
    \centering
    \includegraphics[width=0.8\columnwidth]{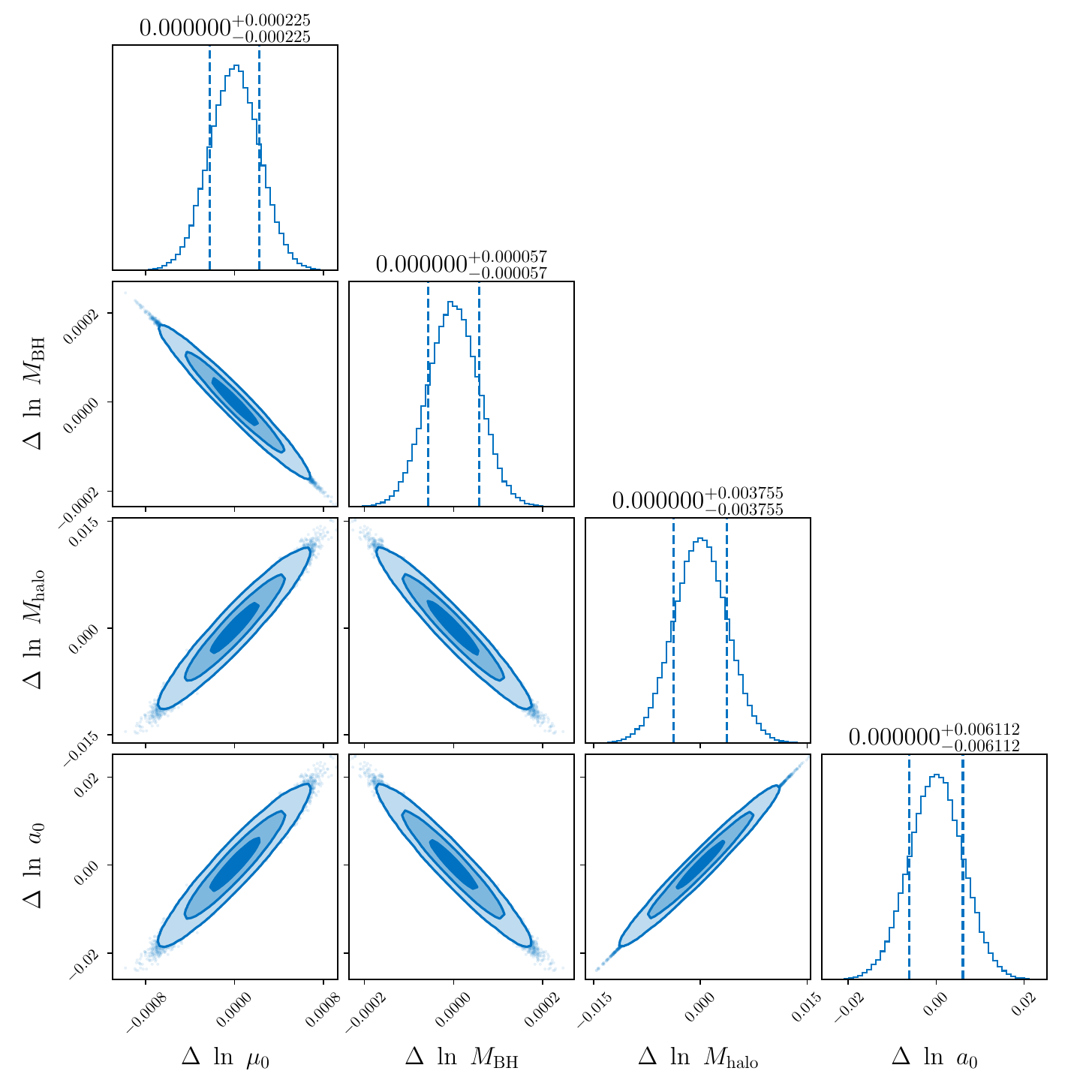}
    \caption{Corner plot for the probability distribution of masses and DM parameters, inferred after EMRI with $(M_{\rm halo}=1~M_{\rm BH},a_0=100~M_{\rm BH})$ and taking into account dynamical friction and accretion. Diagonal boxes refer to marginalized distributions. Vertical lines show the 1-$\sigma$ interval for each waveform parameter. The contours correspond to the $68\%$, $95\%$, and $99\%$ probability confidence intervals.}
    \label{DMACcorner}
\end{figure}
The results of other specific DM configurations are shown in Table \ref{fimtable}.
\begin{table}
 \centering
 \begin{tabular}{|c|c|c|c|c|c|c|}
  \hline
   &\multicolumn{3}{|c|}{DM}&\multicolumn{3}{|c|}{DM+DF+AC}\\
\hline
$(M_{\rm halo},a_0)$  & Mismatch & $\sigma_{ \ln M_{\rm halo}}$  & $\sigma_{ \ln a_0}$  & Mismatch & $\sigma_{ \ln M_{\rm halo}}$ & $\sigma_{ \ln a_0}$ \\ 
\hline
$(1,10^2)$&1.00 & $9.0\times 10^{-2}$&  $2.5\times 10^{-1}$ &1.00& $3.7\times 10^{-3}$& $6.1\times 10^{-3}$\\  \hline
$(10,10^3)$& 0.98 &  $2.1\times 10^{-1}$  &  $4.6\times 10^{-1}$&0.99  &    $2.3\times 10^{-1}$      &  $4.6\times 10^{-1}$\\  \hline
$(100,10^3)$&1.00 &  $1.1\times 10^{-1}$               & $2.0\times 10^{-1}$     &1.00             &    $4.4\times 10^{-3}$     &      $7.9\times 10^{-3}$\\  \hline

\end{tabular}
\caption{The mismatch between vacuum case and DM with/without dynamical friction and accretion. The errors of parameters $M_{\rm halo}$ and $a_0$ with/without dynamical friction and accretion for specific configuration $M_{\rm halo}$ and $a_0$.}\label{fimtable}
\end{table}
The mismatch between vacuum and DM environment is so large that the DM environment should be easily detected.
However, due to correlations, DM parameters cannot be measured with sufficient accuracy unless high-density DM or taking into account the dynamical friction and accretion.
Without dynamical friction and accretion, the effects from DM mainly have the form of $M_{\rm halo}/a_0$ and the polar fluctuations coupled with the fluid may break possible degeneracies so it is possible for us to detect DM parameters $M_{\rm halo}$ and $a_0$ \cite{Cardoso:2022whc}.
When the density of dark matter $M_{\rm halo}/a_0^2$ is larger than $10^{-5}$, including dynamical friction and accretion can improve the accuracy of the measurements of DM parameters $M_{\rm halo}$ and $a_0$.
The reason is that the effects of dynamical friction and accretion depending on the density of DM which has the form of $M_{\rm halo}/a_0^2$ can break the degeneracies deeper.

\section{Conclusions}\label{sec5}
We assess the detector's ability to detect whether an EMRI is immersed in a dark matter halo and to measure its characteristic spatial scale $a_0$ and mass $M_{\rm halo}$.
By using the generic, fully relativistic formalism to study gravitational wave emission by extreme mass ratio systems in spherically symmetric established by Cardoso et al \cite{Cardoso:2022whc}, we calculate the energy emissions and GWs from EMRIs surrounded by dark matter.
We find that the total energy flux emitted by EMRIs is not only dependent on the DM compactness $M_{\rm halo}/a_0$ but also sensitive to the DM density $M_{\rm halo}/a_0^2$.
So they can break possible degeneracies to enable us to detect DM parameters $M_{\rm halo}$ and $a_0$.
We computed the difference between the number of GW cycles accumulated by EMRIs and the mismatch with and without dark matter in circular orbits over one year before the ISCO.
Our results indicate that dark matter can be easily detected when the DM halo compactness parameter $M_{\rm halo}/a_0<10^{-2}$ no matter the density of dark matter $M_{\rm halo}/a_0^2$.
Furthermore, we need a more accurate estimation on the detectability of dark matter because the method of dephasing and mismatch does not take into account the degeneracy of parameters and the correlation between parameters, like the detectability of the secondary spin \cite{Piovano:2021iwv}.
We apply the Fisher information matrix method to estimate the uncertainty of DM parameters.
Due to correlations, DM parameters cannot be measured with sufficient accuracy unless high-density DM or taking into account the dynamical friction and accretion.
The effect of dynamic friction and accretion depends on the density of dark matter.
When the density of dark matter $M_{\rm halo}/a_0^2$ is larger than $10^{-5}$, including dynamical friction and accretion can break possible degeneracies between parameters $a_0$ and $M_{\rm halo}$ and decrease the uncertainty by about one order of magnitude.

\begin{acknowledgments}
This work makes use of the Black Hole Perturbation Toolkit package.
This work was supported by the China Postdoctoral Science Foundation  2023M742297.
\end{acknowledgments}

\appendix
\section{Numerical setup for Fisher information matrix}\label{FIMsetup}
\begin{figure}
    \centering
    \includegraphics[width=0.9\columnwidth]{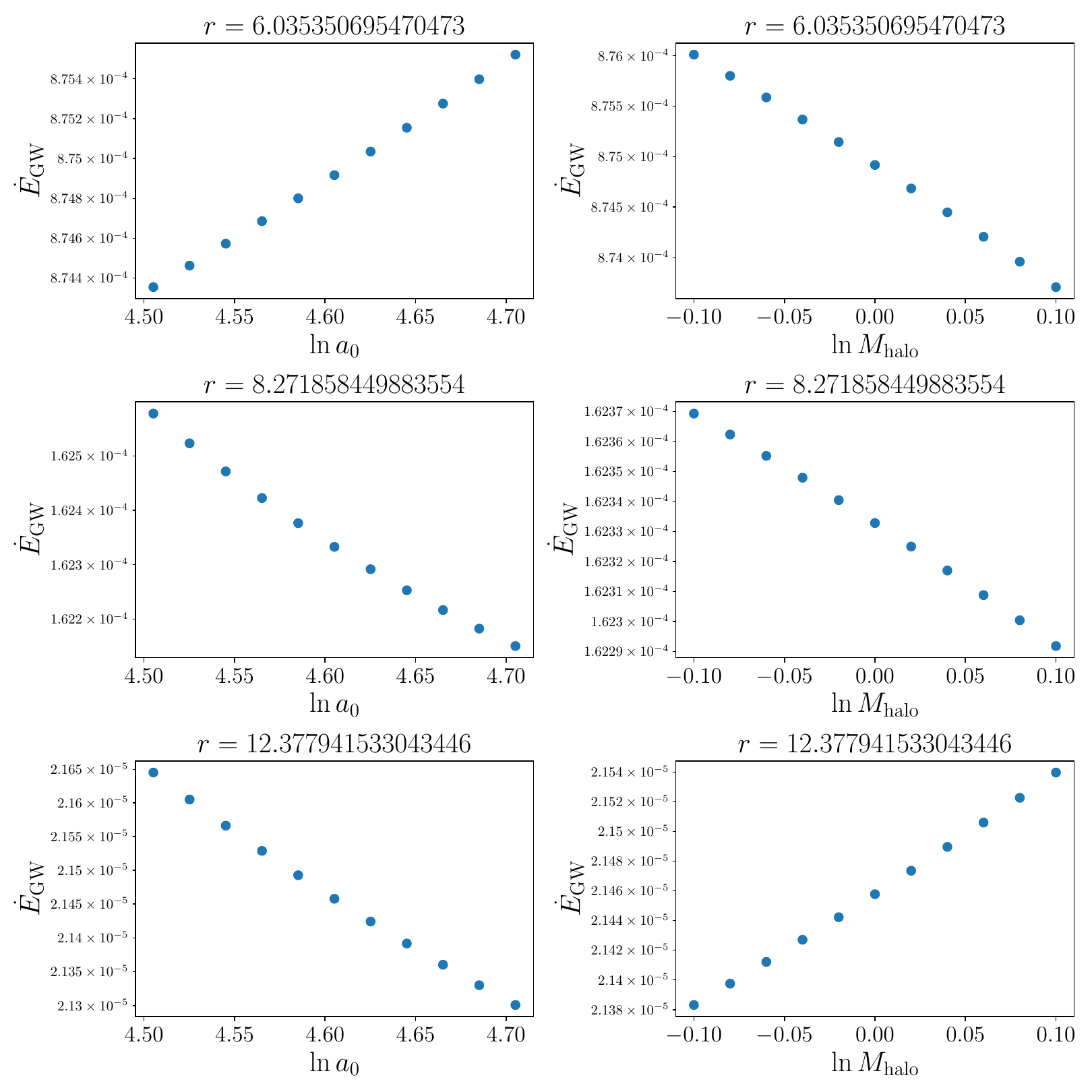}
    \caption{The total GW energy fluxes versus the dark matter parameters $M_{\rm halo}$ and $a_0$ for given orbital radius. The central dark matter parameters values are $M_{\rm halo}=1~M_{\rm BH}$ and $ a_0=10^3~M_{\rm BH}$.}
    \label{PDFLUX}
\end{figure}

In this section, we shall provide technical details on the method and accuracy of the numerical calculations we performed.
We take the dark matter parameters $M_{\rm halo}=1~M_{\rm BH}$ and $ a_0=10^3~M_{\rm BH}$ as a representative of calculating FIM in our discussion and the analysis can be easily extended to other configuration.
In order to perform derivatives we have computed the GW energy loss $\dot{E}_{\rm GW}$ in uniform steps of $\Delta \ln~{M_{\rm halo}}=0.02$ and  $\Delta \ln~{a_0}=0.02$ symmetrically around the central value $M_{\rm halo}$ and $a_0$. 
Figure \ref{PDFLUX} shows the total GW energy fluxes varying with one dark matter parameter $M_{\rm halo}(a_0)$ fixed another dark matter parameter $a_0(M_{\rm halo})$ at the different radial coordinate.
In this way, we can compute the GW energy flux differentiation with respect to the dark matter $M_{\rm halo}$ and $a_0$ at any radial coordinate.
The overall GW energy flux differentiation needs to be numerically interpolated only along the radial coordinate.
We computed derivatives for the Fisher matrix with analytic expression.
Taking parameter $a_0$ for example, the signal can be written in analytic form 
\begin{equation}
h(t)=h(t,r(t),\mu(t),\varphi(t),\xi).
\end{equation}
We can get 
\begin{equation}
\partial_{\ln a_0} h(t)=\frac{\partial h}{\partial r(t)} \frac{\partial r(t)}{\partial \ln a_0}+\frac{\partial h}{\partial \mu(t)} \frac{\partial \mu(t)}{\partial \ln a_0}+\frac{\partial h}{\partial \varphi(t)} \frac{\partial \varphi(t)}{\partial \ln a_0}+\frac{\partial h}{\partial \ln a_0}.
\end{equation}
So all we need is the calculate the orbital evolution derivatives with respect to the dark matter $a_0$.
We can write the orbital evolution equation in a compact form
\begin{equation}
 \dot{\vec y}=\vec F (\vec y,t,\xi), 
\end{equation}
where $\vec y=(r(t),\mu(t),\varphi(t))$.
Now the solutions $\vec y$ will in general be differentiable functions of both $t$ and $\xi$, so we have
\begin{equation}
\dot{\vec y}=\vec F (\vec y (t,\xi),t,\xi), 
\end{equation}
since $\vec y(t,\xi)$ is a function of $(t,\xi)$, we can rewrite the formal orbital evolution equation in the form
\begin{equation}\label{eq2}
\frac{\partial \vec y(t,\xi)}{\partial t}  =  \vec F (\vec y (t,\xi),t,\xi).
\end{equation}
We can differentiate Eq. \eqref{eq2} with respect to $a_0$
\begin{equation}
\frac{\partial}{\partial a_0}    \frac{\partial \vec y(t,\xi)}{\partial t} =\frac{\partial}{\partial a_0} \vec F (\vec y (t,\xi),t,\xi).
\end{equation}
Using the chain rule, we can get 
\begin{equation}\label{eq3}
\frac{\partial}{\partial t} \frac{\partial \vec y(t,\xi)}{\partial a_0}=\partial y_i \vec F (\vec y (t,\xi),t,\xi) \frac{\partial y_i}{\partial a_0}+ \frac{\partial \vec F(\vec y, t,\xi)}{\partial a_0}
\end{equation} 
By solving Eq.~\eqref{eq3}, we can get the orbital evolution derivatives with respect to the dark matter $a_0$, then get the signal differentiation with respect to $a_0$ in the analytical form for calculating FIM.
The waveform $h(t)$ in the time domain is sampled with the step $\Delta t=30$ seconds.


\providecommand{\href}[2]{#2}\begingroup\raggedright\endgroup

\end{document}